\documentclass[aps,prb,twocolumn,superscriptaddress,notitlepage]{revtex4-1}

\usepackage{epsfig}
\usepackage{amssymb} 
\usepackage{amsfonts} 
\usepackage{mathtools} 
\usepackage{dcolumn} 
\usepackage{graphicx} 
\usepackage{color}  
\usepackage{bm}  
\usepackage{comment} 
\usepackage{units}
\usepackage{sidecap} 
\usepackage{textcomp} 

\usepackage{array} 

\newcommand{\bpm}{\begin{pmatrix}}
\newcommand{\epm}{\end{pmatrix}}
\newcommand{\be}{\begin{eqnarray}}
\newcommand{\ee}{\end{eqnarray}}
\newcommand{\ba}{\begin{array}}
\newcommand{\ea}{\end{array}}

\def \tl{Tl$_2$Ba$_2$CuO$_{6+\delta}$}
\def \ybco{YBa$_2$Cu$_3$O$_{6+\delta}$}
\def \hbco{HgBa$_2$CuO$_{6+\delta}$}
\def \lsco{La$_{2-x}$Sr$_x$CuO$_4$}

\begin{document}
\title{Studying angle-dependent magnetoresistance oscillations of cuprate superconductors in a model with antiferromagnetic reconstruction and magnetic breakdown}

\author{Sylvia K. Lewin}
\affiliation{Department of Physics, University of California, Berkeley, California 94720, USA}
\affiliation{Materials Sciences Division, Lawrence Berkeley National Laboratory, Berkeley, California 94720, USA}

\author{James G. Analytis}
\affiliation{Department of Physics, University of California, Berkeley, California 94720, USA}
\affiliation{Materials Sciences Division, Lawrence Berkeley National Laboratory, Berkeley, California 94720, USA}

\date{\today}

\begin{abstract}
We calculate angle-dependent magnetoresistance oscillations (AMRO) for interlayer transport of cuprate superconductors in the presence of ($\pi,\pi$) order. The order reconstructs the Fermi surface, creating magnetic breakdown junctions; we show how such magnetic breakdown effects can be incorporated into calculations of interlayer conductivity for this system.  We successfully fit experimental data with our model, and these fits suggest a connection between ($\pi,\pi$) order and the anisotropic scattering observed in overdoped cuprates. This work paves the way for the use of AMRO as a tool to distinguish different kinds of ordered states.
\end{abstract}

\maketitle

\section{Introduction}

Understanding the nature of broken symmetry phases in the thermodynamic phase diagram of the cuprates is a key step toward understanding the origin of high-temperature superconductivity.  For example, the discovery of the pseudogap \cite{Tallon2001} has fueled the search for many kinds of order \cite{Shekhter2013,Xia2008,He2011,mcelroy_coincidence_2005,he_single-band_2011, neto_ubiquitous_2014, cyr-choiniere_onset_2015}, including nematic phases that could strongly enhance $T_c$ \cite{Lederer2015}.  Yet of the broken symmetries connected to unconventional superconductivity, antiferromagnetism remains one of the most important, appearing in cuprate, iron-pnictide, organic, and heavy fermion materials \cite{Monthoux2007,sachdev_antiferromagnetism_2012,davis_concepts_2013}.

Unambiguous evidence of the presence of Fermi surface reconstruction arising from broken symmetry order has come from quantum oscillation measurements in both hole-doped \cite{Doiron-Leyraud2007,Sebastian2012,Ramshaw2015} and electron-doped \cite{Helm2009} cuprates at low temperatures and high magnetic fields, but the nature of the broken symmetry remains a matter of considerable debate. Antiferromagnetic ($\pi,\pi$) reconstruction has been proposed for both the hole- and electron-doped materials \cite{harrison_cuprate_2007,Helm2009}, but recent evidence for a (possibly field-induced) charge density wave \cite{Wu2011,Wu2013,Croft2014,leboeuf_thermodynamic_2013} has suggested more complex orders are driving the reconstruction.

The ability to experimentally differentiate between these different ordered states is crucial. In this work, we suggest that interlayer angle-dependent magnetoresistance oscillations (AMRO) can be used to distinguish different kinds of long-range ordered states in the cuprates. Angle-dependent magnetoresistance is a sensitive probe of the Fermi surface of a material \cite{hussey_coherent_2003,kartsovnik_high_2004,lebed_magic_2004,yamaji_angle_1989,goddard_angle-dependent_2004} and can therefore be used to investigate the geometry of a reconstructed Fermi surface.  The measurement is also sensitive to the energy scale of any (translational) symmetry-breaking order \cite{Shoenberg}. This energy scale is related to a ``magnetic breakdown field," as we will describe below. Importantly, these effects on AMRO can be observed even in materials that do not show quantum oscillations.  Thus, the measurement is useful in systems in which sample disorder is high, or in which the order has a small correlation length \cite{mcelroy_coincidence_2005,neto_ubiquitous_2014}. 

AMRO data from \tl\, provided the earliest transport evidence for the existence of a three-dimensional Fermi surface in an overdoped cuprate \cite{Hussey2003}. The temperature evolution of the AMRO is consistent with a superposition of isotropic and anisotropic scattering rates about the Fermi surface \cite{Abdel-Jawad2006a}, and it was determined that these do not have the same temperature dependence: the isotropic scattering rate is quadratic with temperature (as expected of an ordinary Fermi liquid) while the anisotropic scattering rate is linear (connecting it to the non-Fermi liquid physics of the cuprate phase diagram). Additionally, the anisotropic scattering is strongest in the anti-nodal region of the Fermi surface. Therefore, it has been suggested that the anomalous scattering temperature dependence may be related to ($\pi,\pi$) fluctuations\cite{Abdel-Jawad2006a, Kokalj2012, dellanna_electrical_2007}, possibly originating from antiferromagnetism. We show that the effects of such order on the AMRO can indeed make a natural connection with the physics of the observed scattering anisotropy, even though there is no static order in the system.

We demonstrate this connection by simulating the AMRO of a model cuprate material in the presence of antiferromagnetic order.  The interpretation of AMRO measurements requires efficient and versatile calculations of the magnetotransport of a given Fermi surface so that models can be compared to experimental results.  These calculations are more challenging in the presence of static order that reconstructs the Fermi surface.  We have developed a general method to perform such calculations for quasi-two-dimensional (Q2D) materials, based on previous work in organic metals \cite{Nowojewski2008}.  It is both easy to implement and computationally inexpensive. In Section \ref{sec:model} we use this method to calculate the interlayer magnetoresistance of a tetragonal Q2D material in the presence of ($\pi,\pi$) order and including the effects of magnetic breakdown.  In Section \ref{sec:tl} we apply this model to the known Fermi surface of \tl~\cite{Hussey2003} and show that the temperature dependence of the AMRO can be captured by this magnetic breakdown model. In Section \ref{sec:discuss} we discuss the physical consequences of this model and its potential range of applicability for distinguishing between different kinds of broken symmetry order in the cuprates.  Our general method is laid out in detail in Appendix \ref{app:method}.

\section{AMRO in the presence of ($\pi,\pi$) order}
\label{sec:model}

\begin{figure}[h!]
    	\includegraphics[width=0.48\textwidth]{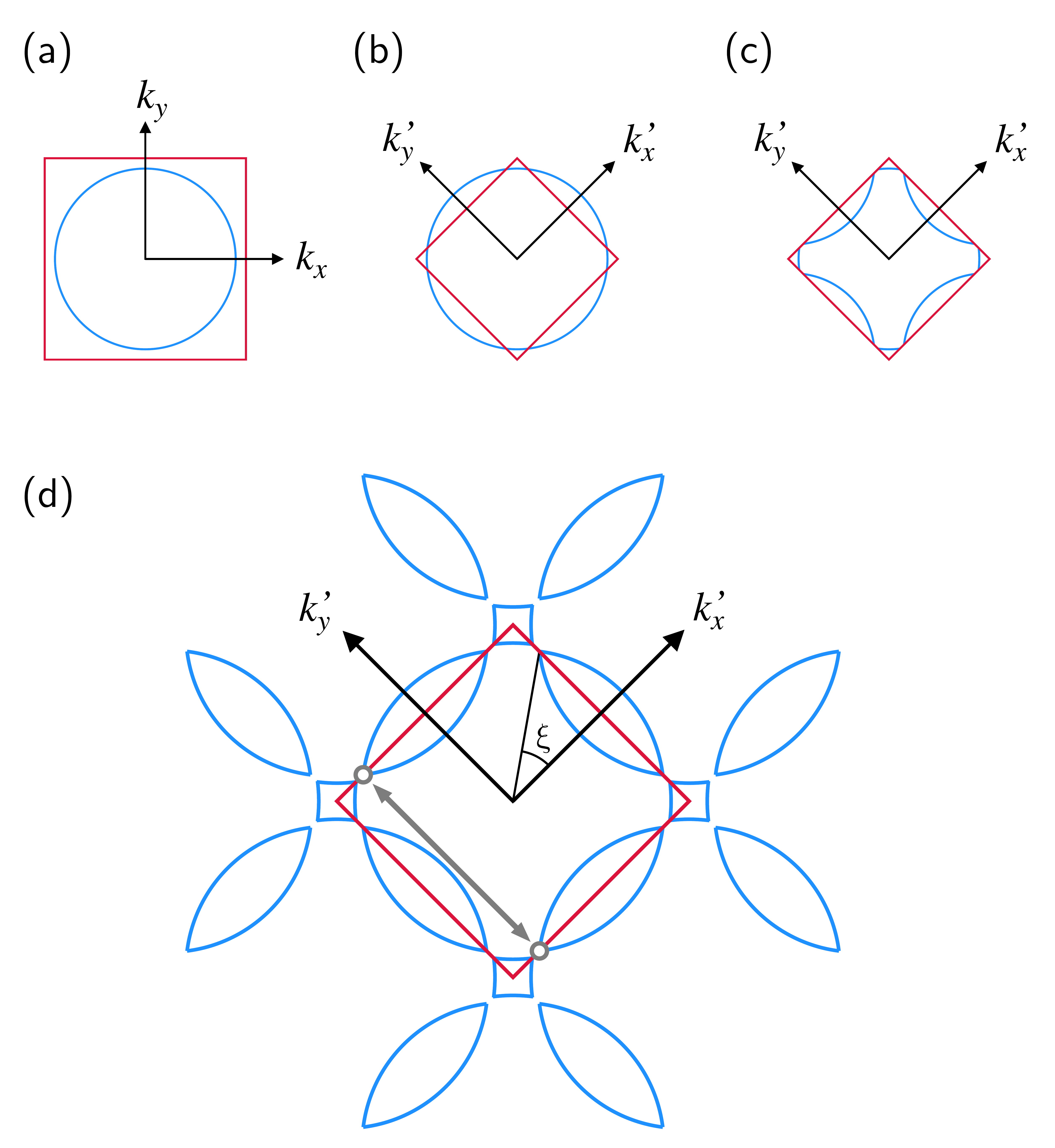}
  	\caption{(Color online) Fermi surface reconstruction of a Q2D material under ($\pi$,$\pi$) antiferromagnetic order, as viewed along $k_z$.  (a) The Fermi surface (FS) and first Brillouin zone (BZ) of a Q2D tetragonal material; (b) the reconstructed BZ and (c) reconstructed FS of the material following the onset of ($\pi,\pi$) antiferromagnetic order; (d) the repeated-zone view of the reconstructed FS, illustrating the small cross-sections of FS that have replaced the unreconstructed cylindrical FS.  The grey line illustrates Bragg diffraction between two magnetic breakdown junctions at the BZ boundary.  The angle $\xi$ is also defined here; it will be used in our conductivity calculations.}
  	\label{fig:FSrec}
	\end{figure}

As a first application of our method, we wish to understand how the AMRO of the cuprates is affected by antiferromagnetism.  We therefore consider the case of a Q2D tetragonal material under static ($\pi,\pi$) antiferromagnetic order (though the model below can be applied to any ($\pi,\pi$) order).  As shown in Figure \ref{fig:FSrec}(a), the original Brillouin zone of such a material will have a square cross-section with primitive reciprocal lattice vectors along $k_x$ and $k_y$; we define all azimuthal angles in this paper with respect to $k_x$.  In the presence of $(\pi,\pi)$ antiferromagnetic order, the Brillouin zone is halved in cross-section, resulting in a reconstruction of the Fermi surface as shown in Figure \ref{fig:FSrec}(b,c).  This new reconstructed Brillouin zone will have primitive reciprocal lattice vectors along $k_x'$ and $k_y'$, which are rotated by 45 \textdegree with respect to $k_x$ and $k_y$.

Quasiparticles traversing the Fermi surface will Bragg diffract at the reconstructed Brillouin zone boundaries, so they will travel along three distinct Fermi surface pockets as shown in Figure \ref{fig:FSrec}(d).  However, in a large magnetic field, the quasiparticle path in real space may be curved sufficiently to avoid Bragg diffraction.  This is known as \textit{magnetic breakdown} (MB), and can be thought of as a tunneling in $k$-space from one pocket to the next \cite{Shoenberg}.  The probability to tunnel in this way is given by $p = e^{-B_0/B}$, where $B_0$ is the breakdown field and is a material-dependent constant proportional to the gap in $k$-space between Fermi surface sections \footnote{Note that in a two-dimensional material, the quasiparticle faces a larger $k$-space tunneling barrier when its orbit is tilted; we therefore write $p = e^{-B_0/Bcos(\theta)}$ so that $B_0$ itself has no angular dependence}.  At every instance the quasiparticle path reaches a Brillouin zone boundary, the quasiparticle may either Bragg diffract or undergo MB; thus, such points in the quasiparticle path are known as MB junctions.

We must take the effect of these MB junctions into account when calculating conductivity.  The conductivity of a Q2D material in a magnetic field can be calculated using the Shockley tube integral form of the Boltzmann transport equation \cite{Ziman1972},

	\begin{equation}\label{eq:Shockley2}
	\begin{split}
		\sigma_{\alpha\beta} = \frac{e^2}{4\pi^3 \hbar^2} \frac{m^*}{\omega_c} &\int dk_B \int_0^{2\pi}  					v_\alpha(\varphi_0,k_B) d\varphi_0 \times \\*
		&\int_{\varphi_0}^{\infty}  v_\beta(\varphi,k_B)
			e^{-(\varphi - \varphi_0)/\omega_c\tau} d\varphi
	\end{split}
	\end{equation}
where $\varphi_0$ is the initial azimuthal position of the quasiparticle and $\varphi$ is its position after some time $t$ has passed\footnote{We have defined $\varphi \equiv \varphi_0 + \varphi'$ to rewrite the bounds of integration as given by Ziman.  The slight difference between this form and that given by Ziman is then merely the difference of whether one considers the quasiparticle to be traveling clockwise or counterclockwise about the Fermi surface.}. The effective mass of the quasiparticle is represented by $m^{*}$ and the cyclotron frequency is $\omega_c = eB/m^{*}$.  The velocities in Eq. \ref{eq:Shockley2} are Fermi velocities.

\begin{figure}
    	\includegraphics[trim=0.5cm 0.5cm 1.5cm 0.1cm, clip=true,width=0.25\textwidth]{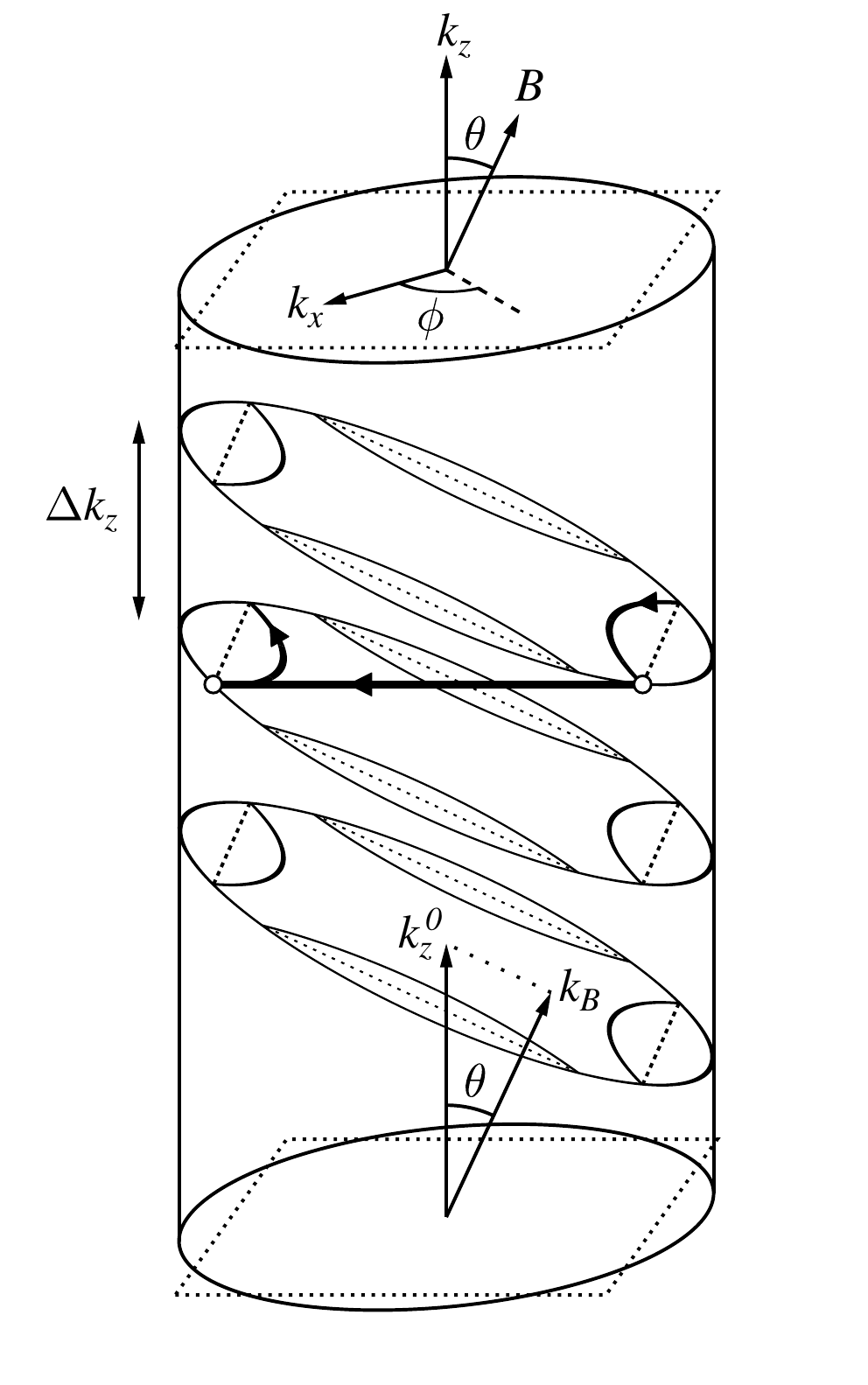}
  	\caption{When a magnetic field is applied to a Q2D material, quasiparticles will trace out orbits on the cylindrical Fermi surface that are perpendicular to the applied field.  On the lowest orbit in this figure, we define the parameters $k_z^0$ and $k_B$.  On the upper two orbits, we illustrate the fact that a quasiparticle undergoing Bragg diffraction moves to a different cross-section of the Fermi surface.  The dashed lines represent the Brillouin zone boundaries for the reconstructed Fermi surface.  Note that the azimuthal angle, $\phi$, is defined with respect to $k_x$ of the \textit{unreconstructed} system.}
  	\label{fig:Bragg}
	\end{figure}

The vector $\bm{k}_B$ points parallel to the magnetic field and defines the orbital path of a quasiparticle.  We integrate across all values of its magnitude.  For a given magnitude, the tip of the vector will touch a single quasiparticle orbit which can be defined by $k_z^0$, the $k_z$-position of the orbital plane at the center of the Fermi surface (see Figure \ref{fig:Bragg}).  The magnetic field's direction is defined by a polar angle ($\theta$) with respect to $k_z$ and an azimuthal angle ($\phi$) with respect to $k_x$.  We can write $|\bm{k}_B | = k_z^0cos(\theta)$ and therefore convert our integral over $\bm{k}_B$ to one over $k_z^0$.

Since AMRO is a probe of interlayer conductivity, we want to calculate $\sigma_{zz}$.  This means we need an expression for $v_z$, which we can obtain from a symmetry-constrained model of the Fermi surface.  The following equation describes a Q2D Fermi surface of a layered tetragonal material with simple cosine warping along $k_z$ \cite{Hussey2003, Analytis2007}:
	\begin{equation}\label{eq:FS}
	E_F(k_z,\varphi) = \frac{\hbar^2 k_F^{\parallel 2}(\varphi)}{2m^*} - \frac{2t_{\perp}a}{\pi}\cos\left(\frac{k_z c}{2}\right)F(\varphi).
	\end{equation}
In the above, $t_{\perp}$ is the interlayer hopping, while \mbox{$k_F^{\parallel}(\varphi)$} and $F(\varphi)$ parameterize the Fermi surface in the azimuthal cylindrical coordinate.  The in-plane and out-of-plane lattice parameters are denoted by $a$ and $c$, respectively. Using $v_z = \frac{1}{\hbar}\frac{dE(\bm{k})}{d k_z},$ we find the interlayer velocity to be
	\begin{equation}
	v_z(\bm{k},\varphi) = \frac{t_{\perp}ac}{\pi \hbar}\sin\left(\frac{k_z c}{2}\right)F(\varphi).
	\end{equation}.

When a quasiparticle Bragg diffracts at the Brillouin zone boundary it will have a momentum change given by a reciprocal lattice vector, so its momentum in the $z$-direction will not change.  The particle will jump to a different ``slice" of the Fermi surface\cite{Nowojewski2008} of $k_z^0$ (see Figure \ref{fig:Bragg}), changing its value but preserving $k_z$.  The amount by which $k_z^0$ changes after Bragg diffraction depends on which MB junctions are involved; for each pair of MB junctions, the value of $\Delta k_z$ can be calculated using purely geometric means (see Appendix \ref{app:deltas}).  Therefore, we can see that for a given quasiparticle,
	\begin{equation}
	k_z(\varphi) = k_z^0 - k_F^{\parallel}(\varphi)\tan(\theta)\cos(\varphi - \phi) +
	\sum\limits_{j} n_j(\varphi)\Delta k_z^{(j)},
	\end{equation}
where $\Delta k_z^{(j)}$ are the possible changes of $k_z$ from Bragg diffraction and $n_j$ are the number of times each has occured \cite{Nowojewski2008}. Note that we have neglected the influence of $t_{\perp}$ on particle motion, which is a reasonable ommission except for $\theta \approx 90$\textdegree. Setting $n_j(\varphi_0) = 0$ we find
	\begin{equation}
	\begin{split}
		\sigma_{zz} = \frac{m^{*}\cos(\theta)}{\omega_c} & \int_{-2\pi/c}^{2\pi/c} dk_z^0 
		\int_0^{2\pi} d\varphi_0\ F(\varphi_0) \sin\left[\frac{ck_z(\varphi_0)}{2}\right] \\*
		& \int_{\varphi_0}^{\infty}  d\varphi\ F(\varphi) \sin\left[\frac{ck_z(\varphi)}{2}\right]
		e^{-(\varphi - \varphi_0)/\omega_c \tau}
	\end{split}
	\end{equation}
up to a constant of proportionality.\\
\ \\
Performing the integration over $k_z^0$, we arrive at
	\begin{equation}
	\begin{split}
	\sigma_{zz} = &\frac{2\pi}{c} \times \frac{m^{*} \cos(\theta)}{\omega_c} \int_0^{2\pi} 
	d\varphi_0\ F(\varphi_0) \int_{\varphi_0}^{\infty} d\varphi\ F(\varphi) \times \\*
	&\cos\left(-G(\varphi) + \frac{c}{2}\sum\nolimits_{j} n_j(\varphi)\Delta k_z^{(j)}  + G(\varphi_0)\right) 
	e^{-(\varphi - \varphi_0)/\omega_c \tau}
	\end{split}
	\end{equation}
where we define $G(\varphi) \equiv \frac{c}{2} \cdot k_F^{\parallel}(\varphi)\tan(\theta)\cos(\varphi - \phi)$.\\
\ \\
We neglect the constant prefactor and use $\cos(x) = \text{Re}[e^{ix}]$ to write
	\begin{equation}
	\begin{split}
\sigma_{zz} = & \frac{m^{*} \cos(\theta)}{\omega_c} \text{Re}\left[ \int_0^{2\pi} d\varphi_0\ F(\varphi_0) \int_{\varphi_0}^{\infty} d\varphi\ F(\varphi) \times \right.\\
&\left. e^{i[G(\varphi) - G(\varphi_0)]}e^{-(\varphi - \varphi_0)/\omega_c \tau}e^{\frac{-i c}{2} \sum\nolimits_{j} n_j(\varphi)\Delta k_z^{(j)}}\right]
	\end{split}
	\end{equation}

Note that the value of the integrand changes whenever the quasiparticle undergoes Bragg diffraction, due to the term $\sum\nolimits_{j} n_j(\varphi)\Delta k_z^{(j)}$.  In order to evaluate the integral, we must be able to account for all possible trajectories of each quasiparticle.

Following the method of Nowojewski \textit{et al.} \cite{Nowojewski2008}, we will separately consider the motion of quasiparticles starting in the 8 different segments of the Fermi surface, then sum their contributions to the conductivity.  To do so, we rewrite the above integral in a vectorized form:

	\begin{equation}
	\sigma_{zz} = \frac{m^{*} \cos(\theta)}{\omega_c} \cdot \text{Re}\left[\bm{\lambda}_{\varphi_0} \cdot (\bm{\lambda}_{\text{init}} + \bm{\Gamma} (\bm{I}-\bm{\Gamma})^{-1}\bm{\lambda}_\varphi) \right]
	\end{equation}

In this equation, the dot product with $\bm{\lambda}_{\varphi_0}$ sums up all the possible initial positions of the quasiparticle, $\bm{\lambda}_{\text{init}}$ describes the initial motion of the quasiparticle up to an MB junction, and $\bm{\lambda_{\varphi}}$ describes the contribution to conductivity when the particle is between MB junctions.  Each vector is 8-dimensional, and they are defined as follows:
	\begin{equation}
	\begin{split}
	\bm{\lambda_{\varphi_0}}[j] &\equiv e^{-\bm{M}_j/\omega_c \tau}  \int_{\bm{M}_j}^{\bm{M}_{j+1}} d\varphi_0 F(\varphi_0) e^{\varphi_0/\omega_c \tau}  e^{-iG(\varphi_0)} \\
	\bm{\lambda_{\varphi}}[j] &\equiv e^{\bm{M}_j/\omega_c \tau}  \int_{\bm{M}_j}^{\bm{M}_{j+1}} d\varphi F(\varphi) e^{-\varphi/\omega_c \tau}  e^{iG(\varphi)}\\
	\bm{\lambda_{\text{init}}}[j] &\equiv e^{\bm{M}_j/\omega_c \tau} \int_{\varphi_0}^{\bm{M}_{j+1}} d\varphi F(\varphi) e^{-\varphi/\omega_c \tau} e^{iG(\varphi)}
	\end{split}
	\end{equation}
where 
	\begin{equation}
	\begin{split}
	&\bm{M} \equiv \frac{\pi}{4}-\xi + \\
&[0,\ \ 2\xi,\ \ \frac{\pi}{2},\ \ \frac{\pi}{2}+2\xi,\ \ \pi,\ \ \pi + 2\xi,\ \ \frac{3\pi}{2}, \ \ \frac{3\pi}{2}+2\xi,\ \ 2\pi]
	\end{split}
	\end{equation}
is a vector giving the azimuthal position of each MB junction and the angle $\xi$ is defined in Figure \ref{fig:FSrec}(d).

The matrix $\bm{\Gamma}$ accounts for the connections between orbit segments, as well as the exponential damping of the integrand upon traversing a segment of Fermi surface.  For our system, it is an $8 \times 8$ matrix:
	\begin{widetext}
	\begin{equation}
	\bm{\Gamma} \equiv 
	\bpm 0 & ap & 0 & 0 & aqe^{-\frac{i}{2} \Delta k_{z}^{(2)}} & 0 & 0 & 0 \\
	0 & 0 & bp & 0 & 0 & 0 & 0 & bqe^{-\frac{i}{2} \Delta k_z^{(3)}} \\
	0 & 0 & 0 & ap & 0 & 0 & aqe^{-\frac{i}{2} \Delta k_z^{(4)}} & 0 \\
	0 & bqe^{-\frac{i}{2} \Delta k_z^{(5)}} & 0 & 0 & bp & 0 & 0 & 0 \\
	aqe^{-\frac{i}{2}\Delta k_z^{(6)}} & 0 & 0 & 0 & 0 & ap & 0 & 0 \\
	0 & 0 & 0 & bqe^{-\frac{i}{2} \Delta k_z^{(7)}} & 0 & 0 & bp & 0 \\
	0 & 0 & aqe^{-\frac{i}{2} \Delta k_z^{(8)}} & 0 & 0 & 0 & 0 & ap \\
	bp & 0 & 0 & 0 & 0 & bqe^{-\frac{i}{2} \Delta k_z^{(1)}} & 0 & 0 \epm
	\end{equation}
	\end{widetext}
where $q = 1-p$ and we have defined $a \equiv e^{-2\xi/\omega_c\tau}$ and $b \equiv e^{-(\pi/2-2\xi)/\omega_c\tau}$.  See Appendix \ref{app:method} for an explanation of the elements of $\bm{\Gamma}$.

As a simplification, we have assumed that the gaps that open in the Fermi surface upon reconstruction are of a negligible length in $k$-space: we take the MB junction that ends one section of the Fermi surface to be in the same position as the MB junction that begins the next section.

With $\sigma_{zz}$ in this vectorized form, we can quickly calculate numerical values for the conductivity with varying $\theta$ and $\phi$.

\section{Application to a cuprate superconductor}
\label{sec:tl}

	\begin{figure*}
    	\includegraphics[trim=0cm 0cm 0cm 0cm, clip=true,width=\textwidth]{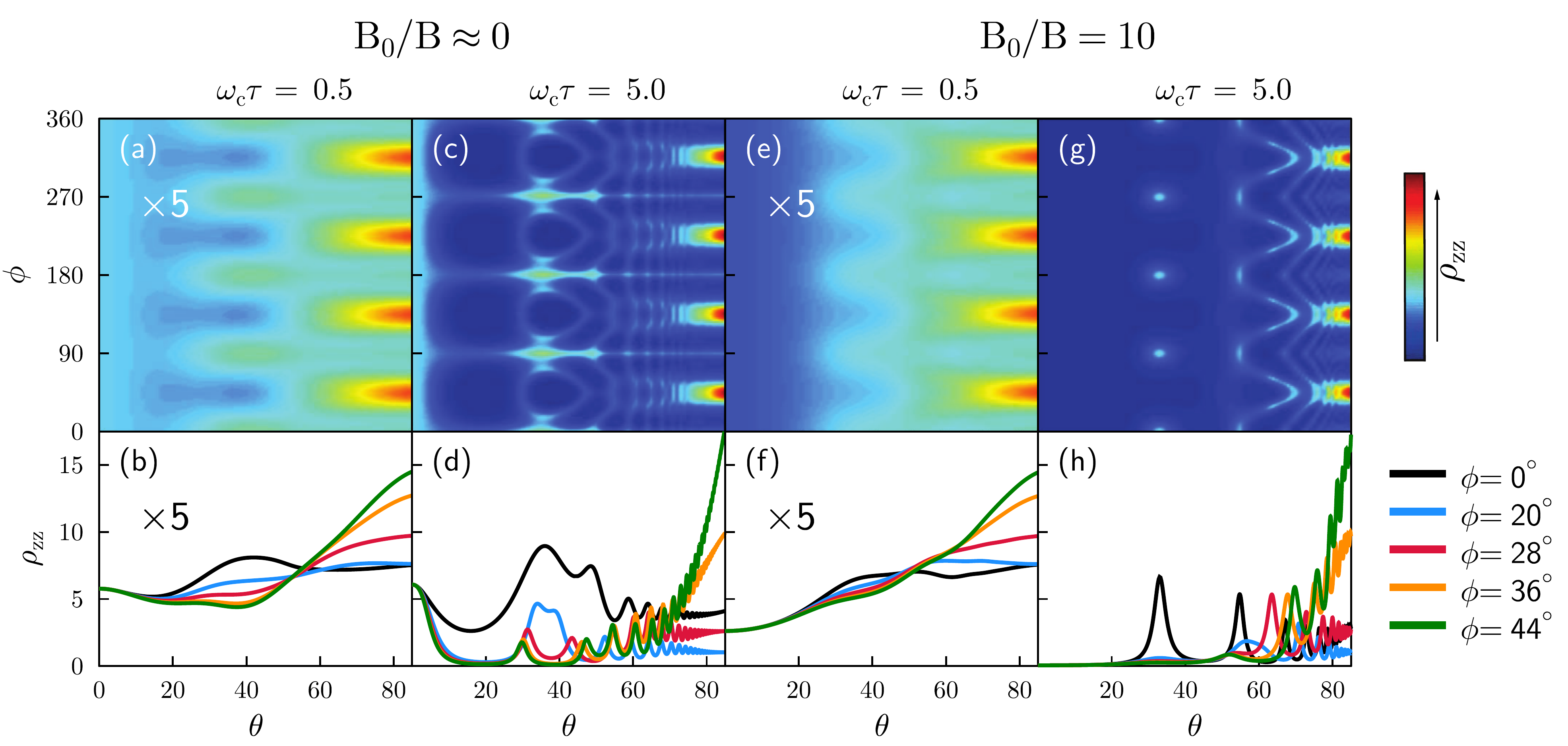}
  	\caption{(Color online)  Calculated dimensionless resistivity of \tl\, as a function of the orientation of applied magnetic field for different values of $\nicefrac{B_0}{B}$ and $\omega_c\tau$.  The resistivity in plots (a,b) and (e,f) is scaled up by a factor of 5 relative to the remaining plots, for visual clarity.  (a,c,e,g) Interlayer resistivity as a function of both $\theta$ and $\phi$.  (b,d,f,h) Interlayer resistivity as a function of $\theta$ for select values of $\phi$; these azimuthal angles were chosen to facilitate comparison to Ref. \onlinecite{Hussey2003}.}
  	\label{fig:AMRO}
	\end{figure*}

We are now in a position to apply this model to a real system. We focus on \tl, since this is the cuprate that has been studied the most with AMRO \cite{Hussey2003, Abdel-Jawad2006a, Analytis2007, French2009}. As described by Hussey {\it et al.}\cite{Hussey2003} the Fermi surface of \tl\ can be parameterized by \mbox{$k_F^{\parallel}(\varphi) \equiv  k_{00} + k_{40}\cos(4\varphi)$} and \mbox{$F(\varphi) \equiv k_{21}\sin(2\varphi) +  k_{61}\sin(6\varphi) + k_{101}\sin(10\varphi)$}.  The coefficients $k_{mn}$ label an expansion of the Fermi surface in cylindrical harmonics appropriate for the space group symmetry of this material \cite{Analytis2007}. 

The AMRO of an unreconstructed Fermi surface can be produced by setting $B_0=0$, as shown in Figure \ref{fig:AMRO} (a-d) for two convenient values of $\omega_c\tau$. Note that $\omega_c\tau=0.5$ reproduces the experimental AMRO observed by Hussey {\it et al.} at 4.2 K \cite{Hussey2003}.  The AMRO for a system with ($\pi,\pi$) antiferromagnetic order is shown in Figure \ref{fig:AMRO} (e-h). This shows many qualitative differences with the unreconstructed state. The peak at $\theta = 0$ is strongly suppressed in the reconstructed Fermi surface.  In addition, there are more Yamaji angles (peaks in the AMRO) for low polar angles $\theta$ in the unreconstructed state than the reconstructed state.

The evolution of the AMRO as we go from $\nicefrac{B_0}{B} \approx 0$ to $\nicefrac{B_0}{B} = 10$ for $\omega_c\tau = 0.5$ bears a striking resemblance to the evolution of the AMRO in \tl\ with increasing temperature, most notably the disappearance of the hump at $\theta = 0$ \cite{Abdel-Jawad2006, Abdel-Jawad2007}.  This seems surprising given that \tl\,is not known to exhibit any static antiferromagnetic order, though it has been shown to have strong antiferromagnetic fluctuations \cite{LeTacon2013}.

\begin{figure}
    	\includegraphics[trim=0.3cm 0.3cm -0.7cm 0.2cm, clip=true,width=0.48\textwidth]{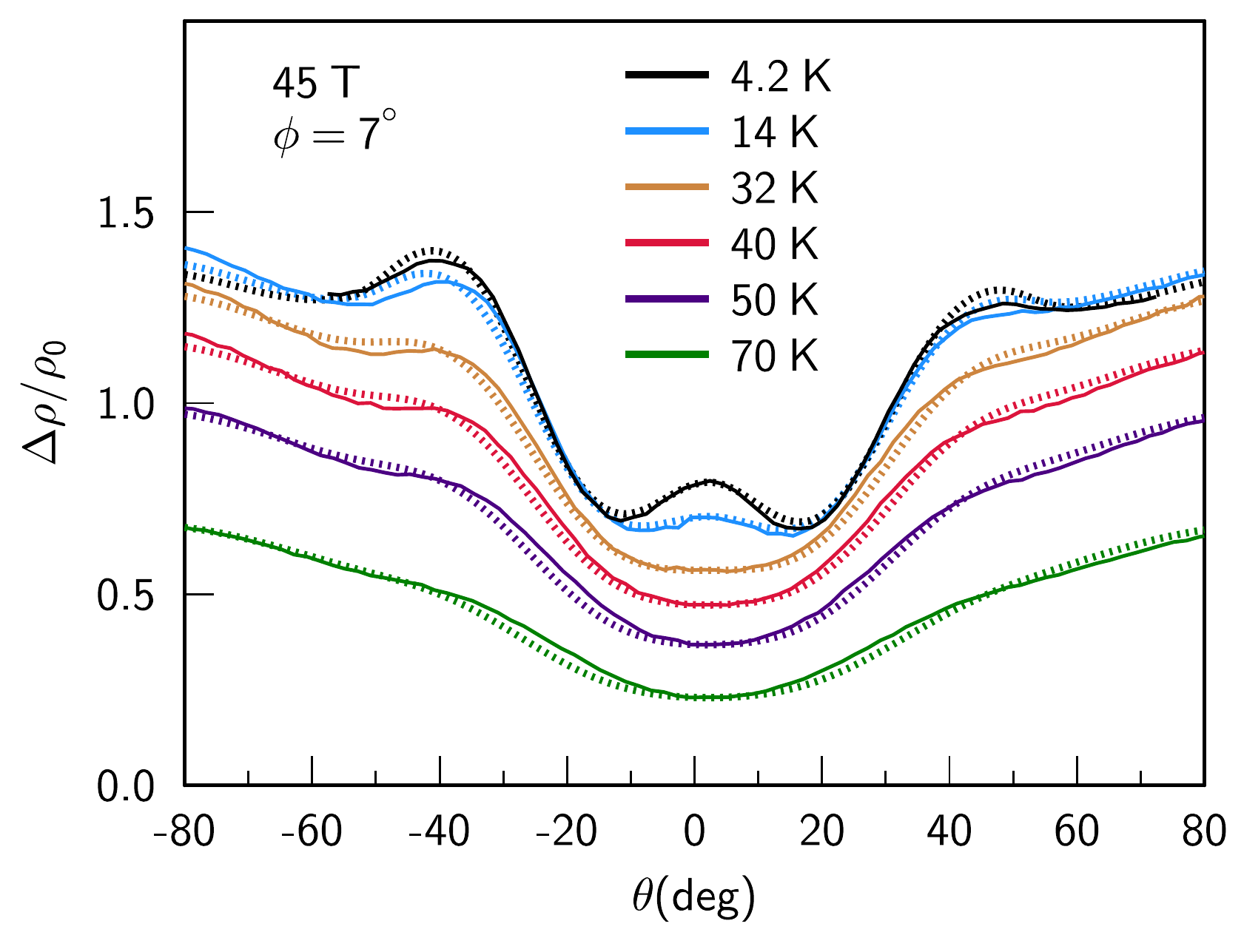}
  	\caption{(Color online) Temperature dependence of the interlayer AMRO of overdoped \tl\, ($T_c =$ 15 K) at a fixed field of 45 T and a fixed azimuthal direction of $\phi = 7$\textdegree.  The solid lines are c-axis magnetoresistivity data, taken from French \textit{et al.} \cite{French2009}.  These data have been normalized to the zero-field resistivity of the sample at each temperature.  The dashed lines are simulations of AMRO for \tl\, under antiferromagnetic order, calculated as described in the text.  When producing these simulations, all the parameters related to Fermi surface geometry were fixed and the only parameters allowed to vary with temperature were $\omega_c\tau$ and $B_0$.}
  	\label{fig:fit}
	\end{figure}

\begin{figure}
    	\includegraphics[trim=-0.2cm 0.3cm -0.2cm -0.5cm, clip=true,width=0.48\textwidth]{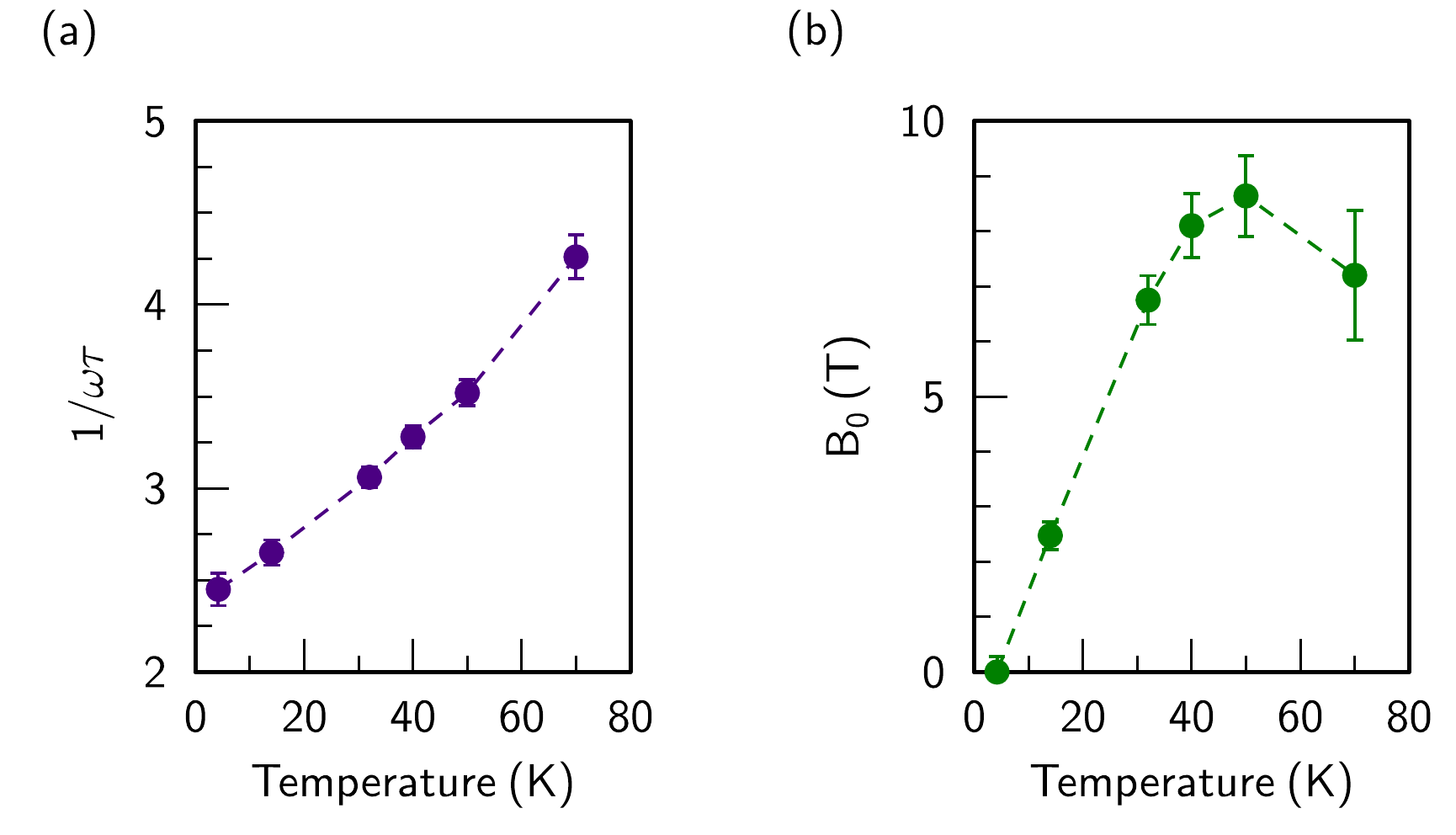}
  	\caption{(Color online)  Temperature dependence of a) $\nicefrac{1}{\omega_c \tau}$ and b) $B_0$ extracted from fits to data.  Error bars are standard errors extracted from the covariance matrix of the least-squares fitting at each temperature (see Appendix \ref{app:params}).}
  	\label{fig:params}
	\end{figure}

We explore the possibility that our AMRO calculations can capture some of the physics of \tl. Using the form of $\sigma_{zz}$ above, we have produced simulations of out-of-plane resistivity as a function of angle using existing data for a sample with $T_c =$ 15 K reported in Ref. \onlinecite{French2009}.  The low-temperature (4.2 K) AMRO of \tl\, is well-fit by a simple model with no antiferromagnetic order ($B_0=0$), and using this data the functions $k_F^{\parallel}(\varphi) $ and $F(\varphi)$ that describe the Fermi surface can be fully determined in good agreement with previous work \cite{Hussey2003, Abdel-Jawad2006, Analytis2007, French2009}.  See Appendix \ref{app:params} for more information on our determination of these parameters.  We used geometric methods to solve for $\xi$ and $\Delta k_z^{(j)}$ in this system, as explained in Appendices \ref{app:xi} and \ref{app:deltas} respectively.  To study the temperature-dependent AMRO above 4.2 K we allowed only two free parameters: $\omega_c \tau$ and $B_0$. Note that in contrast to Ref. \onlinecite{French2009}, $\omega_c\tau$ is fixed to be {\it isotropic} with azimuthal angle $\varphi$.  We ran simulations across a large range of parameter space and used a least-squares fitting approach to determine the values of $\omega_c \tau$ and $B_0$ at each temperature.  Our best fit to the data is shown in Figure \ref{fig:fit}, showing excellent quantitative agreement with the AMRO of \tl. The temperature dependence of $\omega_c \tau$ and $B_0$ can be extracted from these simulations, and these are shown in Figure \ref{fig:params}.  

As can be seen in Figure \ref{fig:AMRO}, more features are apparent in the AMRO when $\omega_c\tau$ is higher, making it easier to distinguish the effects of changing $B_0$.  If $\omega_c \tau$ is decreased (by lowering magnetic fields, raising temperatures, or lowering sample quality), each quasiparticle will traverse less of the Fermi surface before it scatters.  As illustrated in Figure 6 of Ref. \onlinecite{Grigoriev2010}, this causes the amplitude of AMRO to be reduced, which makes an accurate determination of $B_0$ more difficult.  Thus, the error of our fitting parameters is greater at higher temperatures.  Indeed, Ref. \onlinecite{French2009} includes AMRO data taken at 90 K and 110 K, but we were not able to accurately determine $B_0$ at those elevated temperatures.

While Ref. \onlinecite{French2009} reproduced the observed AMRO using an anisotropic scattering rate, we find good quantitative agreement with the data using a magnetic breakdown model with an isotropic scattering rate.  Bragg scattering at a MB junction mimics the effect of an anisotropic scattering rate on an unreconstructed Fermi surface.  However, importantly the magnetic breakdown model connects specific parts of the Fermi surface in a single (Bragg) scattering event, while the anisotropic scattering rate is a broad modulation of the quasiparticle lifetime about the Fermi surface. The similarity of the two models in reproducing the AMRO suggests that the apparent anisotropic scattering rate is a symptom of antiferromagnetism, perhaps involving short-range fluctuations.  This could explain the different temperature dependence of the isotropic and anisotropic components of the scattering rate observed in Ref. \onlinecite{Abdel-Jawad2006a}.

\section{Discussion}
\label{sec:discuss}

The behavior of $\omega_c\tau$ in Figure \ref{fig:params}(a) indicates a natural (approximately linear) increase in the scattering rate with temperature. The evolution of $B_0$ may reflect deeper physics. As shown in Figure \ref{fig:params}(b), $B_0$ increases quickly with temperature, peaking around 45 K. The parameter $B_0$ is a measure of the probability of Bragg scattering.  For a static reconstructed Fermi surface, this is related to the separation between reconstructed sections, which is in turn proportional to the bandgap \cite{Shoenberg}.  Therefore, under static reconstruction we would expect $B_0$ to be largest at 0 K and decrease weakly with increasing temperature \cite{Fenton2005, Shoenberg}.  In the presence of antiferromagnetic fluctuations, similar scattering events might still occur at points where the reconstructed Brillouin zone intersects the Fermi surface. In this case, $B_0$ will play two roles: in addition to parameterizing the separation between sections of Fermi surface, it also reflects the probability of Bragg scattering within the time/length-scale of the fluctuations \footnote{The general form of the breakdown probability would be $p = 1 + g(T)(e^{-B_0/B\rm{cos}(\theta)} + 1)$, where $g(T)$ parameterizes the temperature-dependent strength of fluctuations.  However, adding extra parameters in this way does not add any clarity to our interpretation of the AMRO data.}.  Note that in the overdoped cuprates, there is a known crossover in the transport from Fermi liquid- to non Fermi liquid-like behavior with increasing temperature that is thought to be associated with critical fluctuations \cite{Hussey_dichotomy_2011}.  In this picture, the increase of $B_0$ with temperature (Figure \ref{fig:params}(b)) can be interpreted as an increase in antiferromagnetic fluctuations.  As the temperature rises and antiferromagnetic fluctuations grow, quasiparticles have a non-zero chance of undergoing Bragg diffraction when they reach MB junctions, so $B_0$ attains a non-zero value.  At still higher temperatures, the antiferromagnetic correlation time is so short that the effect of Bragg scattering decreases, resulting in a decrease in $B_0$. The evolution of $B_0$ looks strikingly similar to the evolution of the imaginary part of the dynamic susceptibility Im$\chi$ (which is a measure of the magnetic scattering) seen in a number of neutron experiments in cuprate superconductors; consider, for example, Figure 10 of Ref. \onlinecite{Rossat-Mignod1992}. We therefore suggest that the temperature dependence of $B_0$ in Figure \ref{fig:params}(b) reflects the effect of antiferromagnetic fluctuations on the magnetotransport.

For the above to be plausible, the antiferromagnetic fluctuations of the system should be on a long enough timescale to affect the quasiparticles' motion about the Fermi surface: the timescale of an antiferromagnetic fluctuation should be longer than the time it takes for a quasiparticle to traverse a section of Fermi surface from one MB junction to the next.  The antiferromagnetic fluctuations in \lsco\, near optimal doping have a frequency that is roughly linearly proportional to temperature \cite{Zha1996}.  Taking this as a guide, we estimate that the timescale of an antiferromagnetic fluctuation will be of the order $\tau_{AF} \sim \tfrac{\hbar}{k_B T}$.  Meanwhile, the time for a quasiparticle to cross the smallest section of Fermi surface between two MB junctions is given by $\tau_{QP} \sim \tfrac{1}{\omega_c} \cdot \tfrac{\nicefrac{\pi}{2}-2\xi}{2\pi}$.  Therefore, our condition $\tau_{AF} > \tau_{QP}$ is equivalent to 
	\begin{equation}
	\hbar \omega_c  > \tfrac{\nicefrac{\pi}{2}-2\xi}{2\pi} \cdot k_B T.
	\end{equation}

For this system the requirement is approximately
	\begin{equation}
	\frac{\omega_c}{T}  >  3.6 \times 10^9 s^{-1} K^{-1}.
	\end{equation}
Using $m^* \approx 5m_e$ \cite{Bangura2010}, we find $\omega_c \approx 1.6 \times 10^{12} s^{-1}$ at 45 T.  Therefore, antiferromagnetic fluctuations could be expected to affect quasiparticle motion up to $T\approx$ 400 K, much higher than the temperature regime studied in this paper.

The magnetic breakdown picture of the effect of antiferromagnetic fluctuations on AMRO could be substantially improved by including a more realistic model of the MB junctions in a fluctuating system that includes, for example, a distribution of ordering wavevectors about ($\pi,\pi$) \cite{harrison_cuprate_2007}. Nevertheless, this simple model captures many of the important features observed in the temperature-dependent AMRO without the need for a multi-component scattering with a different nodal and anti-nodal temperature dependence \cite{French2009, Analytis2007, Abdel-Jawad2006a}. Indeed, the success of this simple model indirectly connects this unusual scattering anisotropy to antiferromagnetism, which may be useful for understanding other transport properties (see Appendix \ref{app:inplane}). Moreover, our results suggest there is a potential link between $B_0$ and the dynamic susceptibility Im$\chi$. If this connection can find a sound theoretical basis, it may open the way for the use of AMRO as an experimental probe of magnetic scattering.  

\section{Conclusion}
\label{sec:conc}

We have developed a simple and computationally inexpensive numerical method to calculate AMRO in layered two-dimensional materials with ($\pi,\pi$) antiferromagnetic order.  This model can be applied to both hole- and electron-doped cuprates with an appropriately adjusted Fermi surface parameterization for direct comparison with experimental data. In addition, our numerical method can easily be applied to ordered states other than antiferromagnetism, such as the charge-ordered states recently proposed in underdoped \ybco \cite{ Wu2013, Wu2011,neto_ubiquitous_2014, leboeuf_thermodynamic_2013} and \hbco \cite{Tabis2014}.  We have shown that an antiferromagnetic Fermi surface reconstruction with a temperature-dependent magnetic breakdown field can fit the AMRO of \tl, an overdoped compound with no static order. The agreement between our fits and the AMRO data suggest that the apparent scattering anisotropy observed in these systems\cite{Abdel-Jawad2006a, French2009, Analytis2007} is connected to antiferromagnetic fluctuations, and indeed that the MB field, $B_0$, can potentially be used as an experimental measure of such fluctuations.  This would make AMRO a good complement to scattering probes of fluctuations, such as neutron scattering and resonant inelastic X-ray scattering.  We propose that future AMRO experiments at higher magnetic fields and in materials where Im$\chi$ has been determined independently by neutron scattering would provide an instructive comparison to test the validity of this connection.

\section{Acknowledgements}

We thank Stephen Blundell, Nicholas Breznay, Toni Helm, Ross McKenzie and Andy Schofield for useful discussions.  We acknowledge support from the Laboratory Directed Research and Development Program of Lawrence Berkeley National Laboratory under the US Department of Energy Contract No. DE-AC02-05CH11231.  S.K.L. acknowledges support from the National Science Foundation Graduate Research Fellowship under Grant No. DGE 1106400.  This research used resources of the National Energy Research Scientific Computing Center, a DOE Office of Science User Facility supported by the Office of Science of the U.S. Department of Energy under Contract No. DE-AC02-05CH11231.

\appendix

\section{General method for calculating conductivity with magnetic breakdown}
\label{app:method}

In this Appendix we describe a step-by-step method to calculate $\sigma_{zz}$ in a Q2D material with magnetic breakdown effects.

\begin{enumerate}

\item Consider the full, warped-cylindrical Fermi surface that would exist were the Fermi surface not reconstructed.  Using existing data or theories, determine a likely form of this Fermi surface as a function of $k_z$ and $\varphi$.  This may be exactly fixed or it may contain free parameters to be fitted.

\item Use the Fermi surface to determine $v_\alpha(\bm{k},\varphi)$ and $v_\beta(\bm{k},\varphi)$ for the element $\sigma_{\alpha \beta}$ in question.  Note that $v_x$ and $v_y$ are not simply proportional to $k_x$ and $k_y$ for a noncircular Fermi surface; see the section on in-plane transport below.

\item Insert these velocities into the Boltzmann transport equation as given in Equation 1 of the main text.  Wherever $k_z$ appears in the integrand, replace it with the following function of $\varphi$:
	\begin{equation}
	k_z(\varphi) = k_z^0 - k_F^{\parallel}(\varphi)\tan(\theta)\cos(\varphi - \phi) +
	\sum\limits_{j} n_j(\varphi)\Delta k_z^{(j)}
	\end{equation}

\item Replace the integral over $k_B$ with an integral over $k_z^0$ multiplied by $\cos(\theta)$, and perform the integration over $k_z^0$.  At this point, it should be possible to write the Boltzmann transport equation in the form
	\begin{equation}
	\sigma_{\alpha \beta} =  C_1 \int_0^{2\pi} d\varphi_0\ f_1(\varphi_0)  \int_{\varphi_0}^{\infty}  d\varphi\ f_2(\varphi) e^{C_2 \sum\nolimits_{j} n_j(\varphi)\Delta k_z^{(j)} }
	\end{equation}
where $f_1$ and $f_2$ are functions, and $C_1$ and $C_2$ are constants.  Note that $C_2$ will be zero if $\beta = x \text{ or } y$.

\item Determine geometrically where the Fermi surface will intersect the (reconstructed) Brillouin zone.  These points are the magnetic breakdown junctions.  Write a vector, $\bm{M}$, giving the azimuthal position of each junction and ending at the location of the first junction plus $2\pi$.  Be sure that the definition of $\varphi = 0$ for this vector is consistent with the definition of $\varphi = 0$ for the Fermi surface warping.  The length of $\bm{M}$ will be $n + 1$, where $n$ is the number of MB junctions around the Fermi surface.

\item Define three vectors of length $n$ as follows:
\begin{equation}
	\begin{split}
	\bm{\lambda_{\varphi_0}}[j] &\equiv e^{-\bm{M}_j/\omega_c \tau}  \int_{\bm{M}_j}^{\bm{M}_{j+1}} d\varphi_0\ f_1(\varphi_0)\\
	\bm{\lambda_{\varphi}}[j] &\equiv e^{\bm{M}_j/\omega_c \tau}  \int_{\bm{M}_j}^{\bm{M}_{j+1}} d\varphi\ f_2(\varphi)\\
	\bm{\lambda_{\text{init}}}[j] &\equiv e^{\bm{M}_j/\omega_c \tau} \int_{\varphi_0}^{\bm{M}_{j+1}} d\varphi\ f_2(\varphi)
	\end{split}
	\end{equation}

\item Define the $n \times n$ matrix $\bm{\Gamma}$.  Each row (column) of $\bm{\Gamma}$ corresponds to a specific \textit{section} of the Fermi surface between two MB junctions.  The first row of $\bm{\Gamma}$ corresponds to the section between the first and second MB junctions, as defined in the vector $\bm{M}$; the second row corresponds to the section between the second and third MB junctions; and so on.  The elements in each row are as follows:

\begin{widetext}
	\begin{equation}
	\bm{\Gamma}_{ij} = 
	\begin{cases}
      	0 & \parbox[t]{.6\textwidth}{if section $i$ has no connection to section $j$} \\
	\ \\
      	a_i p & \parbox[t]{.6\textwidth}{if section $i$ is connected to section $j$ through magnetic breakdown\\ (i.e., the quasiparticle on section $i$ goes to section $j$ by following the full cylindrical Fermi surface)}\\
	\ \\
	a_i (1-p)e^{C_2\Delta k_{z}^{(i+1)\rightarrow j}} \text{\ \ \ \ \ } &  \parbox[t]{.6\textwidth}{if section $i$ is connected to section $j$ through Bragg diffraction\\ (i.e., the quasiparticle on section $i$ goes to section $j$ by following the reconstructed Fermi surface)}
    	\end{cases}
	\end{equation}
\end{widetext}

where $p = e^{-B_0/Bcos(\theta)}$ is the magnetic breakdown probability, and $a_i \equiv e^{-(M_{i+1} - M_i)/\omega_c\tau}$.  The term $a_i$ accounts for the damping of our integrand as the quasiparticle traverses the $i^{th}$ section of the Fermi surface.  Note the term $\Delta k_{z}^{(i+1)\rightarrow j}$: after traversing the $i^{th}$ section of the Fermi surface, the quasiparticle would Bragg diffract from the $(i+1)^{th}$ magnetic breakdown junction.  The terms $\Delta k_{z}^{(i+1)\rightarrow j}$ can be calculated as described in the section ``Calculating $\Delta k_z^{(j)}$'' below.

\item  Using the objects defined above, calculate the conductivity for a given direction of the applied field:
	\begin{equation}
	\sigma_{\alpha \beta}(\theta,\phi) =  C_1 \cdot \text{Re}\left[\bm{\lambda}_{\varphi_0} \cdot (\bm{\lambda}_{\text{init}} + \bm{\Gamma} (\bm{I}-\bm{\Gamma})^{-1}\bm{\lambda}_\varphi) \right]
	\end{equation}

Note that the dot product $\bm{\lambda}_{\varphi_0} \cdot \bm{\lambda}_{\text{init}}$ yields a double integral over $\varphi_0$ and $\varphi$ and must be evaluated as such.

\end{enumerate}

\section{Calculating $\xi$}
\label{app:xi}

\begin{figure}[]
    	\includegraphics[trim=0cm 2cm 0cm 2cm, width=0.35\textwidth]{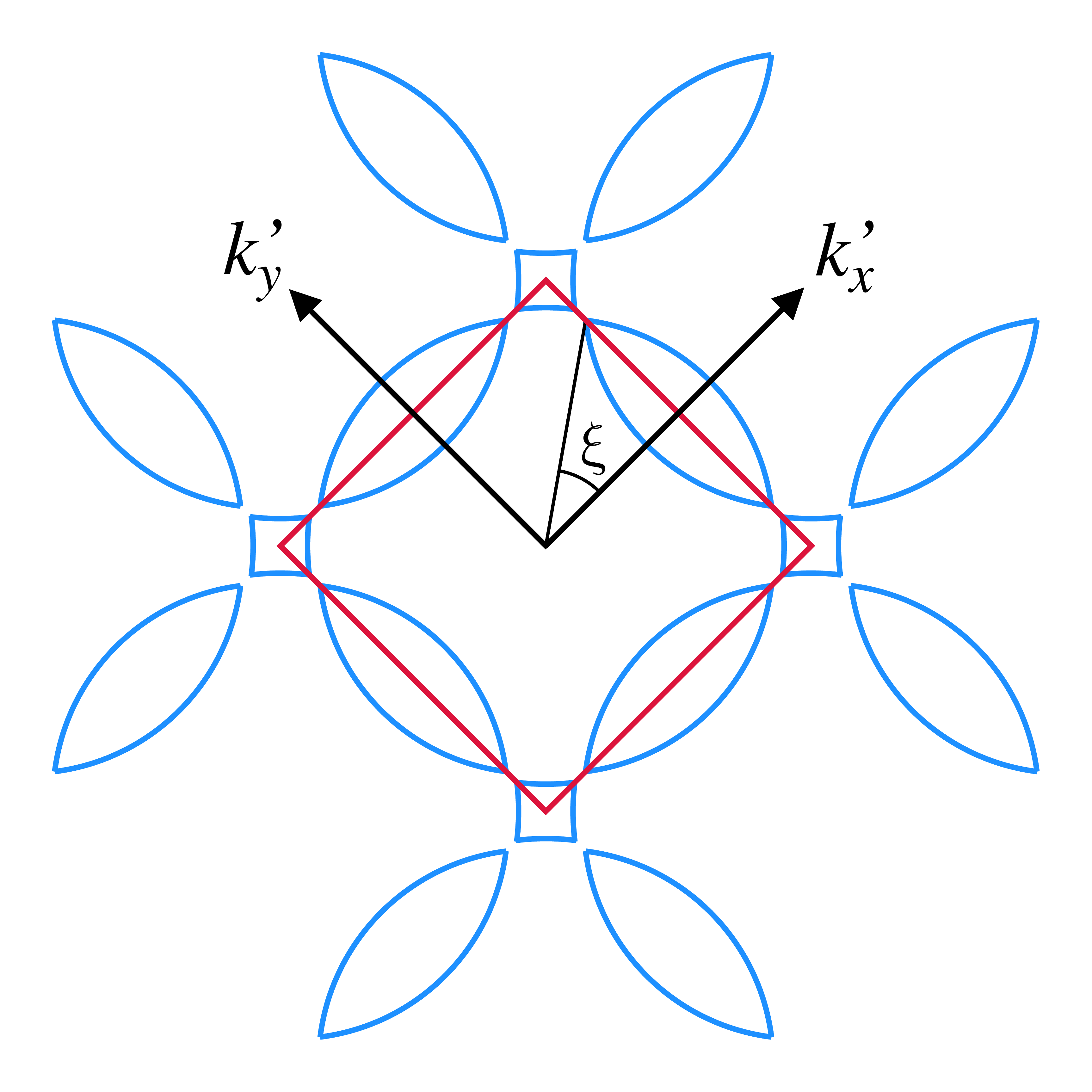}
  	\caption{(Color online) The angle $\xi$ is defined with respect to the reconstructed Fermi surface of \tl.  The Fermi surface is shown in the repeated zone scheme, with the reconstructed Brillouin zone overlaid.}
  	\label{fig:xidef}
	\end{figure}

The angle $\xi$ is defined as shown in Figure \ref{fig:xidef}.  If the Fermi surface were completely cylindrical, it would obey
\begin{equation*}
\cos(\xi) = \frac{\pi}{a_{\text{AF}} k_F}
\end{equation*}
 where $a_{\text{AF}}$ is the in-plane lattice parameter of the antiferromagnetically ordered system and $k_F$ is the Fermi momentum.  We may neglect the interlayer warping of the Fermi surface, which is relatively weak, but not the in-plane warping.  Therefore, we have the relation
\begin{equation*}
\cos(\xi) = \frac{\pi}{a_{AF}(k_{00}+k_{40}\cos(4\xi))}
\end{equation*}
which can be solved self-consistently for $\xi$.  We know that $a_{\text{AF}} = \sqrt{2} a$.  We use $a = 0.3866$ nm, as given by Analytis \textit{et al.}\cite{Analytis2007}.  We use $k_{00} = 7.30\text{ nm}^{-1}$ and $k_{40}  = -0.234\text{ nm}^{-1}$.  These are the values found by French \textit{et al.} from fitting their 4.2 K AMRO data\cite{French2009} and they are consistent with the results of our fits (see above).  Using these values, we find $\xi \approx 40.18$\textdegree.  Due to uncertainty in the Fermi surface fits, we cannot calculate $\xi$ with accuracy beyond two significant digits.  We therefore round to $\xi = 40$\textdegree\ for use in our fits to high-temperature data.

\section{Calculating $\Delta k_z^{(j)}$}
\label{app:deltas}

\begin{figure}[h]
    	\includegraphics[trim=0.5cm 0cm 0.6cm -0.4cm, clip=true[width=0.3\textwidth]{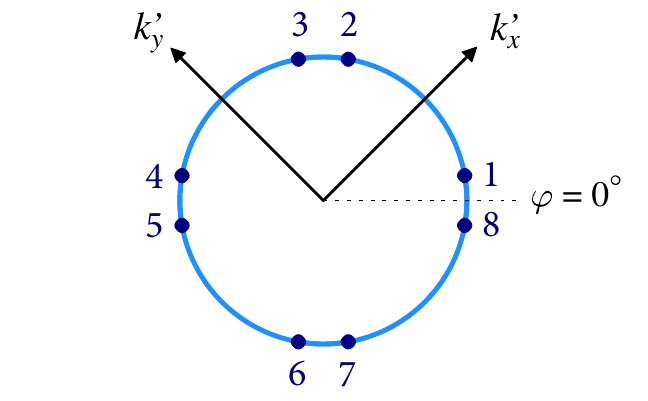}
  	\caption{(Color online) Position of the 8 magnetic breakdown junctions on the Fermi surface of \tl\, under $(\pi, \pi)$ order.  The reciprocal space axes shown correspond to the reconstructed Brillouin zone.}
  	\label{fig:MBjunctions1}
	\end{figure}

As stated in the main text, we can define a vector giving the azimuthal position of each magnetic breakdown (MB) junction as follows:
\begin{equation}
	\bm{M} \equiv \frac{\pi}{4}-\xi + [0,\ \ 2\xi,\ \ \frac{\pi}{2},\ \ \frac{\pi}{2}+2\xi,\ \ \pi,\ \ \pi + 2\xi,\ \ \frac{3\pi}{2}, \ \ \frac{3\pi}{2}+2\xi,\ \ 2\pi]
	\end{equation}
The position of these MB junctions on the (unreconstructed) Fermi surface is shown in Figure \ref{fig:MBjunctions1}.

To find the values of $\Delta k_z^{(j)}$, we must know where a quasiparticle goes when it experiences Bragg diffraction at a given MB junction.  To determine this, we need only see which MB junctions are connected by reciprocal lattice vectors of the reconstructed Brillouin zone.  They are the following: $1 \leftrightarrow 6$, $2 \leftrightarrow 5$, $3 \leftrightarrow 8$, $4 \leftrightarrow 7$.

\begin{figure*}[]
    	\includegraphics[trim=0cm 1.2cm 1cm 1.2cm, width=0.8\textwidth]{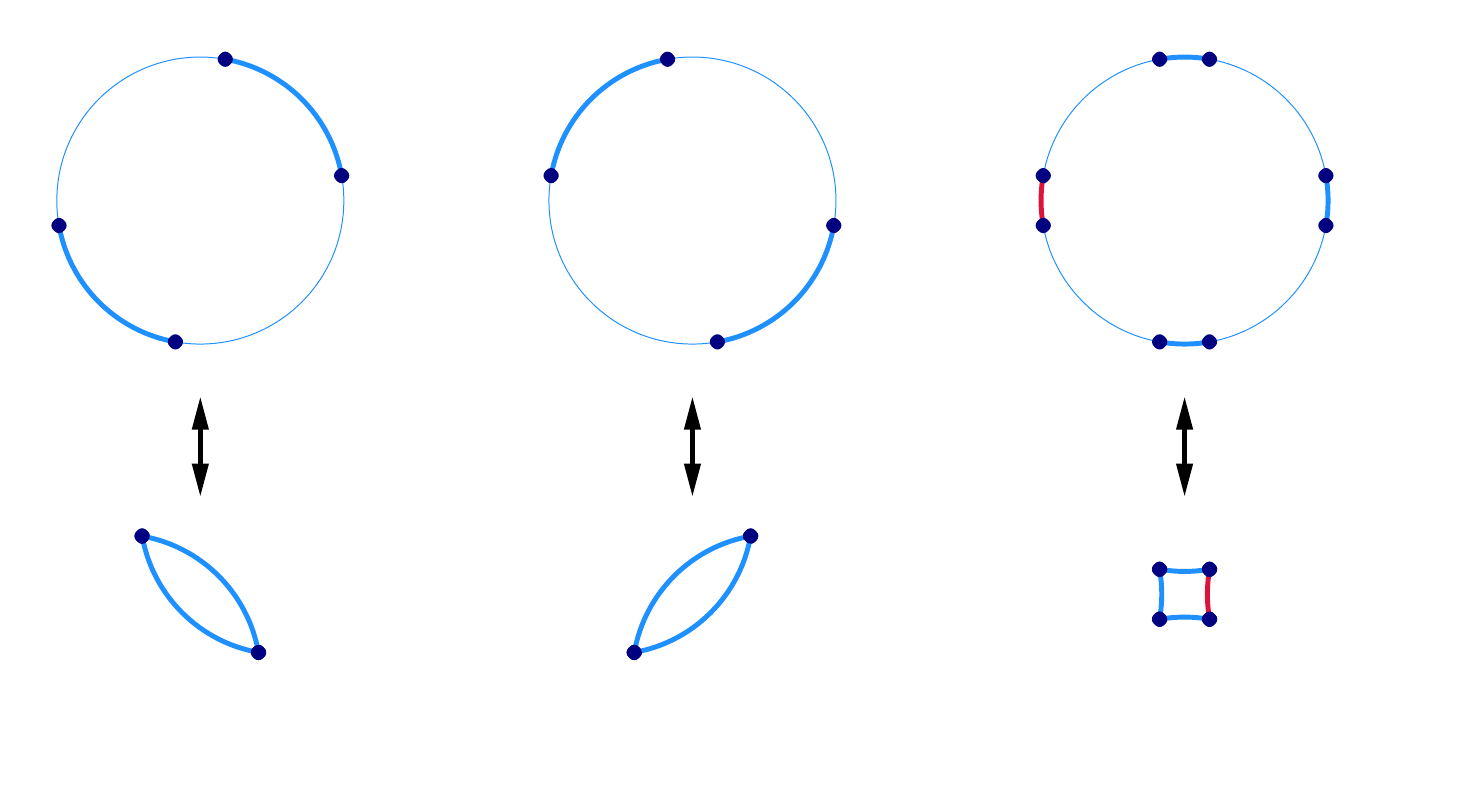}
  	\caption{(Color online) Three small Fermi surface pockets are formed when the Fermi surface of \tl\, is reconstructed under $(\pi, \pi)$ order.}
  	\label{fig:MBjunctions2}
	\end{figure*}

An easy way to understand these pairings is to examine the small Fermi surface orbits that the quasiparticle will follows if it Bragg diffracts at every junction (see Figure \ref{fig:MBjunctions2}).\\
As stated in the main text, a quasiparticle undergoing Bragg diffraction in this system will have $k_z^{(i)} = k_z^{(f)}$.  We neglect the weak interlayer warping of the system; then for a quasiparticle on a particular slice of the Fermi surface, we can write
	\begin{equation}
	k_z(\varphi) = k_z^0 - k_F^{\parallel}(\varphi)\tan(\theta)\cos(\varphi - \phi).
	\end{equation}
This leads to the condition
	\begin{equation}
	\begin{split}
	k_z^{0(i)} - k_F^{\parallel}(\varphi_i)\tan(\theta)\cos(\varphi_i - \phi) = \\
k_z^{0(f)} - k_F^{\parallel}		(\varphi_f)\tan(\theta)\cos(\varphi_f - \phi)
	\end{split}
	\end{equation}
and therefore
	\begin{equation}
	\begin{split}
	\Delta k_z^{(i\rightarrow f)} =  k_z^{0(f)} -  k_z^{0(i)} = & \\ \tan(\theta)[k_F^{\parallel}(\varphi_f)\cos(\varphi_f - &\phi) - k_F^{\parallel}(\varphi_i)\cos(\varphi_i - \phi)].
	\end{split}
	\end{equation}
Since $\varphi_i$ and $\varphi_f$ are given by the vector $\bm{M}$, we now have everything we need to solve for $\Delta k_z^{(i\rightarrow f)}$ for each possible Bragg diffraction.  For example, if a quasiparticle is going from MB junction 1 to MB junction 6 we have $\varphi_i = \frac{\pi}{4}-\xi$ and $\varphi_f = \frac{\pi}{4}-\xi + \pi + 2\xi = \frac{5\pi}{4} + \xi$.  We can use these to solve for $\Delta k_z^{(1\rightarrow 6)}$, which we denote as $\Delta k_z^{(1)}$ for the sake of brevity.

\section{Parameter fitting and error bars}
\label{app:params}

We would expect the parameters $k_{mn}$ to be constant with temperature, as they describe the Fermi surface geometry.  Therefore, we can fit these parameters from our 4.2 K data since we do not expect reconstruction and magnetic breakdown to occur at this temperature.  From band structure calculations \cite{VanderMarel1999} and from previous AMRO studies \cite{Hussey2003}, we expect this material to have no c-axis dispersion along the zone diagonals as well as along the lines $k_x = \pi$ and $k_y = \pi$.  In order for this to be realized, it must be the case that  $1-  \nicefrac{k_{61}}{k_{21}} + \nicefrac{k_{101}}{k_{21}} = 0$ \footnote{Note that only the ratios $\nicefrac{k_{61}}{k_{21}}$ and $\nicefrac{k_{101}}{k_{21}}$ are relevant to our calculations, not the values of these parameters; this is because we are only calculating the interlayer conductivity up to a constant of proportionality, and these parameters do not affect the in-plane conductivity.}.

We simulated conductivity for a wide swath of parameter space and used a least-squares fitting to data to arrive at the following: $k_{00} = 7.34\text{ nm}^{-1}$, $k_{40} = -0.25\text{ nm}^{-1}$, $\nicefrac{k_{61}}{k_{21}} = 0.69$ (and therefore $\nicefrac{k_{101}}{k_{21}} = -0.31$) \footnote{To be precise, we fit for the unitless parameters $k_{00}c$ and $k_{40}c$, then obtained values for $k_{00}$ and $k_{40}$ using $c = 2.32$ nm from Analytis \textit{et al.} \cite{Analytis2007}.}

Simultaneously with fitting the Fermi surface geometry, we used the 4.2 K data to fit for the misalignment of the crystal with respect to the magnetic field; see Analytis \textit{et al.} for details on the significance and calculation of this misalignment \cite{Analytis2007}.  We obtained the best fits to data from $\Phi_{asym} = -0.6$\textdegree, $\Theta_{asym}^x = -2.5$\textdegree, and $\Theta_{asym}^y = -2.8$\textdegree.

Once we have fit these parameters at 4.2 K, the only parameters free to fit for the data as a function of temperature are $B_0$ and $\nicefrac{1}{\omega_c \tau}$.  For each temperature we simulated conductivity across a broad range of $B_0$ and $\nicefrac{1}{\omega_c \tau}$ and used a least-squares fitting to arrive at the following values for the best fits to data:

\begin{table}[h]{}
\begin{tabular}{c | c c}
\ T(K)\ \ & \ $B_0 (T)\  $ & \ $\nicefrac{1}{\omega_c \tau}$\ \ \\[0.5ex] \hline \\[-1.5ex]
4.2 & $0^{\dagger}$  & 2.50 \\[1ex]
14  & 2.5 &  2.65 \\[1ex]
32  & 6.8 & 3.06 \\[1ex]
40  & 8.1 &  3.28 \\[1ex]
50  & 8.6 & 3.52 \\[1ex]
70  & 7.2 & 4.26  \\[0ex]
\end{tabular}
\\ \ \\ $^{\dagger}$We assumed $B_0 = 0$ at 4.2 K in order to perform our fits for Fermi surface geometry and alignment.
\end{table}

The error bars shown on $B_0$ and $\nicefrac{1}{\omega_c \tau}$ in the main text are the standard error of those parameters.  At each temperature, the values of $B_0$ and $\nicefrac{1}{\omega_c \tau}$ that give the best fit to data are those for which the sum of squared error (SSE) between data and simulation is minimized.  We can fit the SSE to a functional form in terms of $B_0$ and $\nicefrac{1}{\omega_c \tau}$ about that minimum.  We use this functional form to approximate the Hessian matrix for these two parameters, the inverse of which is the covariance matrix, $\bm{C}$.  The standard error for each parameter is then simply given by $\sqrt{\nicefrac{\bm{C}_{ii}}{(N-2)}}$, where $N$ is the number of data-points we used for the fitting at that temperature (and 2 is the number of parameters we fit).  Which diagonal element of $\bm{C}$ corresponds to each parameter depends on how we construct the Hessian matrix.

\section{In-plane transport simulations}	
\label{app:inplane}

In addition to calculating $\sigma_{zz}$, we can use the same methods as detailed above to calculate the in-plane components of the conductivity tensor \cite{Nowojewski2010}.  Neglecting the weak interlayer warping of our system, we find \mbox{$v_x(\varphi)= \frac{\hbar}{m^*}k_F^{\parallel}(\varphi)\cos(\varphi-\gamma)$} and  \mbox{$v_y(\varphi)= \frac{\hbar}{m^*}k_F^{\parallel}(\varphi)\sin(\varphi-\gamma)$}.  Here $\gamma$ is the angle between $v_F$ and a vector pointing radially outward towards the Fermi surface, and it is given by
\begin{equation}
\gamma(\varphi) = \tan^{-1}\left[\frac{\partial}{\partial \varphi}(\log k_F(\varphi))\right]
\end{equation}
as described in Ref. \onlinecite{Hussey2003a}.  The procedure is then nearly identical to that for $\sigma_{zz}$, though slightly simplified by the fact that the $\Delta k_z^{(j)}$ terms are not involved in the in-plane calculations.  We can calculate in-plane conductivity exactly, whereas we can only calculate $\sigma_{zz}$ up to a constant of proportionality since we do not know the value of $t_{\perp}$.

Rather than calculating the in-plane transport terms and fitting them to experimental data, we want to see what predictions we can make for in-plane transport based on our analysis of the interlayer transport.  We fit the points from Figure \ref{fig:params} in the main text to analytical functions: a second-order polynomial in temperature for $\nicefrac{1}{\omega_c\tau}$, and a function of the form $\frac{c_1}{T}e^{-c_2/T}$ for $B_0$, as we expect that at higher temperatures $B_0$ must decrease due to weakening antiferromagnetic correlations.
 
\begin{figure*}[]
	\includegraphics[trim=0cm 2cm 0cm 5cm, width=\textwidth]{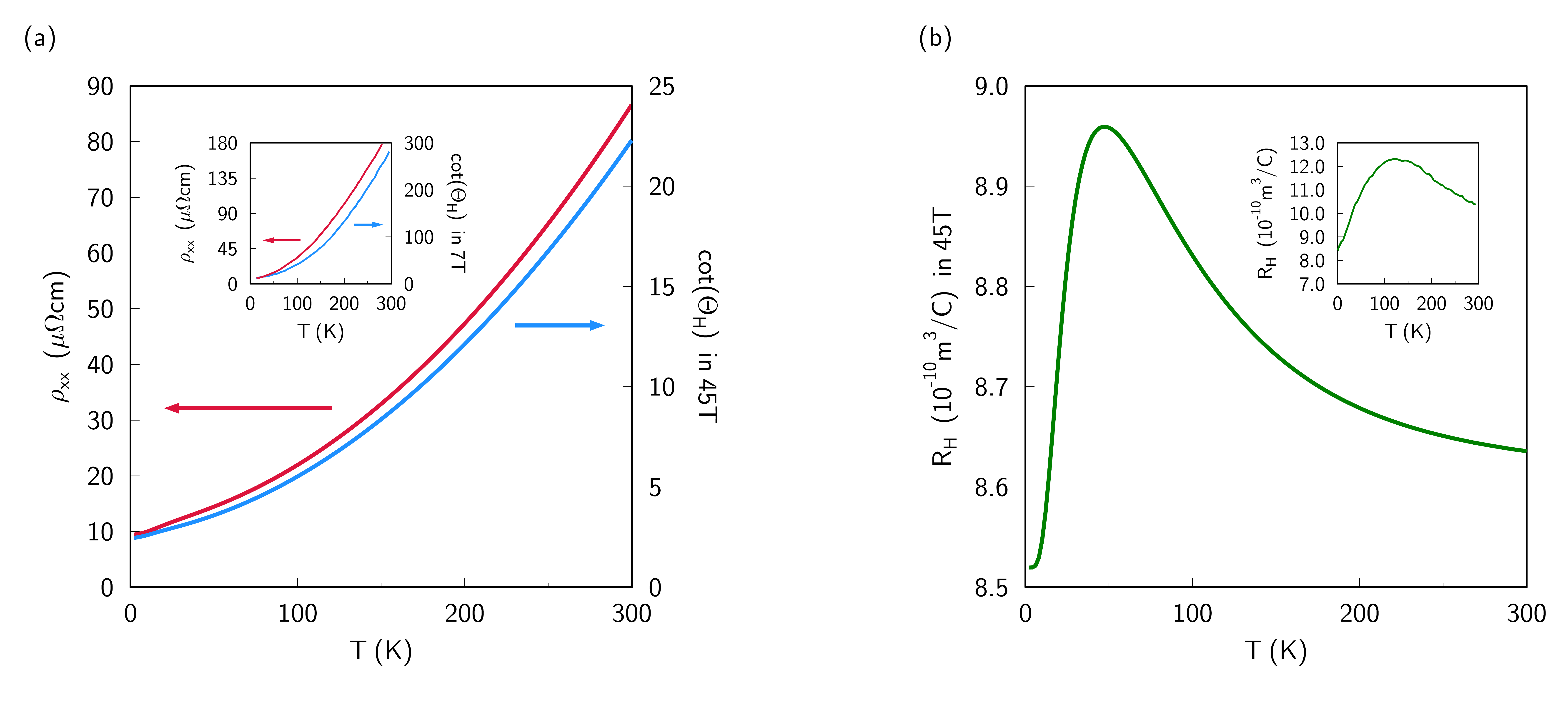}
	\caption{(Color online)  Simulations of in-plane transport for \tl\, compared to experimental data taken from Ref. \onlinecite{Mackenzie1996}.  a) Calculated values for the in-plane resistivity and for the cotangent of the Hall angle in 45 T field plotted versus temperature (inset: data at 7 T).  b) Calculated values of the Hall coefficient as a function of temperature in 45 T field (inset: data at unspecified field).}
	\label{fig:inplane}
	\end{figure*}

Using these analytical functions of our temperature-dependent parameters, we are able to calculate the in-plane transport of \tl\, at any temperature--though such calculations should be interpreted with care because we are extrapolating to higher temperatures using information that comes from 50 K and below.  We can compare these calculations to data taken from comparable samples by Mackenzie \textit{et al.} \cite{Mackenzie1996}, as shown in Figure \ref{fig:inplane}.  Note that the data presented in these figures come from a sample with $T_c$ of 15 K, the same critical temperature as the sample whose AMRO data we have analyzed.

Our simulations of in-plane transport are qualitatively similar to experimental data, though they do not agree quantitatively, especially the Hall angle and Hall coefficient.  It is important to note that in the magnetic breakdown model, where $\nicefrac{B_0}{B}$ plays an important role, we do not have Drude-like resistivity: $\rho_{xy}$ is not directly proportional to the magnetic field.  Therefore, we would have to use Hall data taken at 45 T to truly make a meaningful comparison.  We cannot simply lower the magnetic field strength in our calcuations to match the field at which data was taken, as we only have information on $B_0$ for a 45 T field.   It has been proposed that antiferromagnetism in the cuprates is enhanced by an applied magnetic field \cite{kee_nematicity_2009,franz_magnetic_2002} and therefore we cannot assume that the value of $B_0$ at lower fields matches that at 45 T.

We show these results not because they definitively support or contradict the magnetic breakdown model, but merely in the spirit of sharing the results of our explorations.  Given that we do not know the dependence of $B_0$ on $B$, it seems unlikely that such in-plane calculations can yield strong evidence for or against the suggested model.

\end{document}